\documentclass[12pt]{article}
\usepackage{setspace}
\usepackage{graphicx}
\usepackage{dcolumn}
\usepackage{bm}

\begin{document}

\bibliographystyle{unsrt}

\title{Broadband Purcell effect: Radiative decay engineering with metamaterials}
\author{Zubin Jacob* \\
\small Department of Electrical and Computer Engineering, \\
\small University of Alberta, Edmonton, AB T6G 2V4, Canada \\ 
\small \emph{*zjacob@ualberta.ca} \\ \\
Igor I. Smolyaninov \\
\small Department of Electrical and Computer Engineering, \\
\small University of Maryland, College Park, MD 20742, U.S.A. \\ \\
Evgenii E. Narimanov \\
\small Birck Nanotechnology Center  \\
\small School of Electrical and Computer Engineering \\
\small Purdue University, West Lafayette, IN 47906, U.S.A.}

\date{}

\maketitle
\doublespacing
\begin{abstract}
We show that metamaterials with hyperbolic dispersion support a large number of electromagnetic states that can couple to quantum emitters leading to a broadband purcell effect.  The proposed approach of radiative decay engineering, useful for applications such as single photon sources, fluorescence imaging, biosensing and single molecule detection, also opens up the possibility of using hyperbolic metamaterials to probe the quantum electrodynamic properties of atoms and artificial atoms such as quantum dots.
\end{abstract}


Microcavities and photonic crystals are among the most promising systems for enhancing spontaneous emission by the Purcell effect \cite{purcell1946spontaneous} and to simultaneously collect the emitted photon in a given quantum state \cite{lodahl2004controlling,lounis2005single-photon}. They form the test bed for cavity quantum electrodynamics experiments and aid the major advances in single photon sources . 
However, the high quality of the resonance required for the cavity Purcell effect immediately puts a restriction on the spectral width of the emitter and hence on the possible compatible sources. For example, the reduced linewidth of quantum dots at low temperatures which is ideally compatible with a microcavity for the demonstration of the Purcell effect, is  too wide at room temperatures --- making such systems unviable for Purcell enhancement \cite{baba2003low-threshold}. Moreover, other quantum emitters such as molecules and nitrogen vacancy centers in diamond have broad bandwidth emission not ideally compatible with cavity technology -- and therefore an alternative, non-cavity based approach is needed  \cite{lounis2005single-photon}.
The resulting recent interest towards systems which show a broadband Purcell effect \cite{quan2009broadband,lund2008experimental,jun2009broadband,vesseur2010broadband} opened up the route to a number of new applications -- from  broadband single photon sources \cite{patterson2009broadband,esteban2010optical} to strong coupling of  emitters to plasmons \cite{akimov2007generation,oulton2009plasmon}. In particular, we proposed \cite{jacob2009broadband}  that the large number of electromagnetic states of a hyperbolic metamaterial lead to a divergence in the photonic density of states allowing broadband control over light matter interaction at room temperature. As opposed to conventional methods based on closed cavity Purcell enhancement of spontaneous emission \cite{lounis2005single-photon} or open cavity systems based on photonic crystal waveguides \cite{hughes2004enhanced}, our approach relies on a radiative decay engineering approach using metamaterials \cite{yao2009ultrahigh}, engineering the dielectric repsonse of the medium surrounding the emitter to provide new electromagnetic states for in-coupling. This effect has been demonstrated experimentally \cite{jacob2010engineering,noginov2010controlling}, and substantial enhancement was observed in the spontaneous emission rates of organic dyes and quantum dots when positioned near the surface of a metamaterial with hyperbolic dispersion. In this paper, we expand on our original theoretical prediction of this effect \cite{jacob2009broadband} and present a quantitivate description of the 
spontaneous emission rate enhancement. 

Metamaterials with hyperbolic dispersion, also known as indefinite media \cite{smith2004negative}  lie at the heart of devices such as the hyperlens \cite{jacob2006optical, salandrino2006far-field} and non-magnetic negative index waveguides  \cite{hoffman2007negative}. In an isotropic medium, the dispersion relation
$\frac{k^2}{\epsilon}   =  \frac{\omega^2}{c^2}$ defines a spherical iso-frequency surface in the $k$-space (see Fig.1(a)), thus placing
 an upper cut-off for the wavenumber -- so that high wavevector modes simply decay away. In contrast to this behavior, a strongly anisotropic metamaterial where the the components of the  dielectric permittivity tensor have opposite signs in two ortogonal directions   can support bulk propagating waves with unbounded wavevectors. This can be most clearly seen in the case of uniaxial anisotropy ($\epsilon_z \equiv \epsilon_\parallel$, $\epsilon_x = \epsilon_y \equiv \epsilon_\perp$) where the dispersion relation for the extraordinary (TM-polarized) waves
$
 \frac{k_\parallel^2}{\epsilon_\perp} + \frac{k_\perp^2}{\epsilon_\perp}  =  \frac{\omega^2}{c^2}
$ for $\epsilon_\parallel \epsilon_\perp < 0$ describes a hyperboloid of revolution around the symmetry axis $z$ ( see Fig. 1(b) ) and thus does not limit the magnitude of the wavenumber. 
 Such high-$k$ propagating modes allow for  subwavelength imaging 
 \cite{liu2007far-field} and subdiffraction
mode confinement \cite{govyadinov2006metamaterial}. As we demonstrate in the present paper, these
high wavenumber spatial modes in hyperbolic metamaterials also have strong effect on  quantum-optical phenomena. In particular, they lead to a substantial enhancement of the spontaneous emission, without the need for coupling to a slow waveguide mode or a counter propagating wave as in a resonator. As we will show below, this strong effect also circumvents the need for three dimensional confinement of the emitter \cite{quan2009broadband,lund2008experimental,jun2009broadband,vesseur2010broadband} to achieve the Purcell effect.

In the spirit of the Fermi's golden rule, an increased  number of radiative decay channels due to the high-$k$ states in hyperbolic media, available for an excited atom ensures enhanced spontaneous emission. This can increase the quantum yield by overcoming emission into competing non-radiative
decay routes such as phonons.  A decrease in lifetime, high quantum yield and good collection efficiency can lead to extraction of single photons reliably at a high repetition rate from isolated emitters \cite{lounis2005single-photon}.

We consider the classic example of radiative decay engineering using a substrate which interacts with an emitter placed above it \cite{drexhage1970influence,barnes1998fluorescence}. The spontaneous emission rate in this geometry can be obtained by a straightforward generalization of the standard semiclassical theory \cite{ford1984electromagnetic}. In Fig. 2(a) we plot the corresponding emission decay relaxation time as a function of the distance to the sample, for a hyperbolic metamaterial with the dielectric permittivity tensor $\epsilon_x=1.2+0.1i,\epsilon_y=1.2+0.1i,\epsilon_z=-4+0.1i$. In agreement with the qualitative arguments above, in the close vicinity of the substrate  the availability of the large number of photonic states causes the photons to be preferentially emitted into the metamaterial and the lifetime decreases considerably. Even though the emitter is placed in vacuum and is coupled to the quasi continuum of vacuum states, the large number of states in the metamaterial leads to a Purcell effect without the need for confinement. 

The available radiative channels for the spontaneous  photon emission consist of the propagating waves in vacuum, the plasmon on the metamaterial substrate and the the continuum of high wavevector waves which are evanescent in vacuum but propagating within the metamaterial. The corresponding decay rate into the metamaterial modes when the emitter is at a distance $ a < d \ll \lambda $(where $a$ is the metamaterial
patterning scale)  is 
$
\Gamma^{meta} \approx \frac{\mu^{2}}{8\hbar d^{3}}\frac{2\sqrt{\epsilon_{x}|\epsilon_{z}|}}{(1+\epsilon_{x}|\epsilon_{z}|)}
$

In the close vicinity of the hyperbolic metamaterial, the power from the dipole is completely concentrated in the large spatial wavevector channels (Fig 2(a) inset).The same evanescent wave spectrum when incident on a lossy metal or dielectric would be completely absorbed, causing a non-radiative decrease in the lifetime of an emitter (quenching). On the contrary, the metamaterial converts the evanescent waves to propagating and the absorption thus affects the outcoupling efficiency of the emitted photons due to a finite propagation length in the metamaterial.

Along with the reduction in lifetime and high efficiency of emission into the metamaterial, another key feature of the hyperbolic media  is the directional nature of light propagation.  Fig. 2(b) shows the field along a plane perpendicular to the metamaterial-vacuum interface exhibiting the beamlike radiation from a point dipole. This is advantageous from the point of view of collection efficiency of light since the spontaneous emitted photons lie within a cone \cite{felsen1994radiation}. The group velocity vectors in the medium which point in the direction of the Poynting vector are simply normals to the dispersion curve [Fig. 1]. For vacuum, these normals point in all directions and hence the spontaneous emission is isotropic in nature. In contrast to this behavior, the hyperbolic dispersion medium allows wavevectors only within a narrow region defined by the asymptotes of the hyperbola. Hence the group velocity vectors lie within the resonance cone giving rise to a directional spontaneously emitted photon propagating within the metamaterial. The beamlike nature of the photon in the metamaterial arising solely due to the hyperbolic dispersion has to be distinguished from that obtained by the mode properties of a resonant structure such as a micropost microcavity \cite{lounis2005single-photon} or that of a guided mode in a photonic crystal waveguide \cite{hughes2004enhanced,lecamp2007very}.

The actual realization of the hyperbolic metamaterial introduces deviations from the effective medium description \cite{kidwai2011dipole}. Here we consider a practical realization of a hyperbolic metamaterial consisting of alternate layers of silver ($\epsilon_{Ag}=-2.4+0.48i$) and alumina ($\epsilon_{{\rm Al}_2{\rm O}_3}=2.7$) at a wavelength of $\lambda = 365$ nm. The system consists of 8 layers, each of thickness $ a = 8$ nm   which is easily achievable by current fabrication techniques. We  compute and compare the propagating wave spectrum which is routinely used in ellipsometric measurements for extraction of effective medium parameters. Fig 3(a) shows the plane wave reflection and transmission coefficients computed using transfer matrix techniques in the layered realization superimposed on the effective medium prediction.  Since $a \ll \lambda$, effective medium theory holds in a broad bandwidth (Fig 3 (a) inset).
The lifetime of the dipole on top of the layered h-MM at a distance of $d=\lambda/20$ shows a factor of $F_p = 10$ decrease from the free space lifetime . It should be noted that non-radiative contribution to reduced lifetime (also known as quenching due to lossy surface waves) is considerably decreased when using thin layer of metal as building blocks as compared to thick metal \cite{ford1984electromagnetic}. To ascertain the role of metamaterial modes we compare the lifetime with a single period of ${\rm Ag}/{\rm Al}_2{\rm O}_3$. The hyperbolic system has a lifetime lower by a factor of 2 as compared to a thin piece of metal. This decrease in the lifetime is due to the metamaterial states with large wavevectors as shown by the spectrum of power emitted by the dipole in the vicinity of the layered structure [Fig.3(b) inset]. At a distance of $d=\lambda/10$, the efficiency of emission into metamaterial modes (evanescent in air but propagating within the layered structure) is $\eta_{meta} = 77\%$. Note that the layered realization shows transmission of large wavevector states when compared to a high index dielectric (Fig. 3(b)), as expected from the effective medium theory.

 In conclusion, we have shown that the electromagnetic states of a hyperbolic metamaterial lead to a non-resonant Broadband Purcell effect without the need for confined emitters.  The proposed device based on hyperbolic metamaterials is compatible with a wide variety of sources and capable of room temperature operation due to the broad bandwidth enhancement of spontaneous emission and directional photon emission. Our work paves the way for using metamaterials for applications in quantum nanophotonics ranging from single photon sources to fluorescence based sensing \cite{lakowicz2004radiative}.
 
 The work was partially supported by ARO MURI. Z.J. wishes to acknowledge support from NSERC Discovery and CSEE POP.

\clearpage
\begin{figure}
\centering
\scalebox{0.5}{\includegraphics{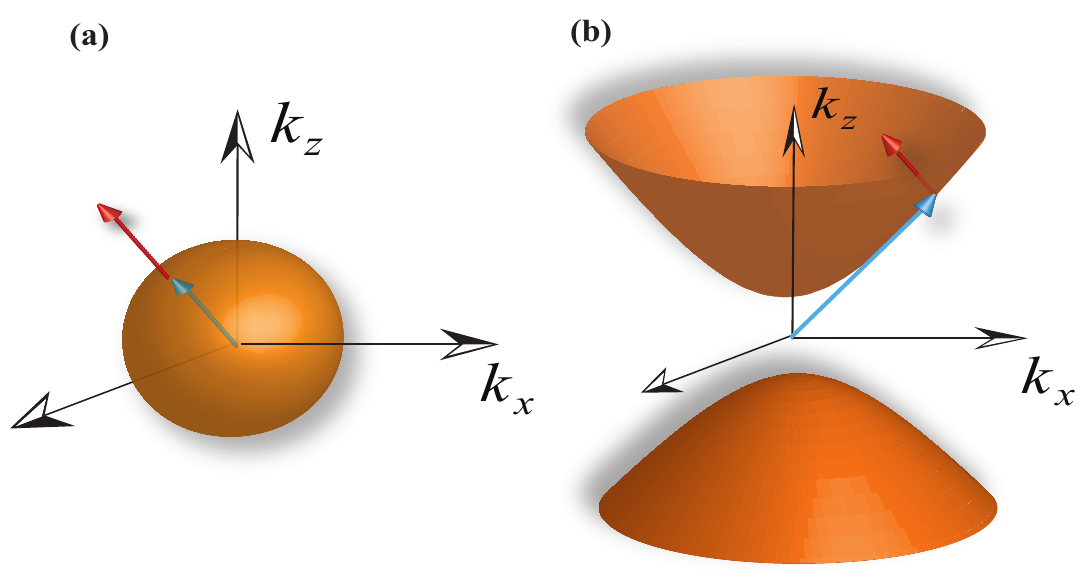}}
\caption{\label{fig1} {\bf a}) Dispersion relation for an isotropic medium. The blue arrow denotes an allowed wavevector, whereas the normal to the dispersion relation gives the direction of the group velocity (red arrow). ({\bf b}) Hyperbolic dispersion relation allowing large number of electromagnetic states with unbounded values of the wavevector (blue arrow). The group velocity vectors (red arrow) lie within a cone which implies light propagation in such media is inherently directional.}
\end{figure}
\clearpage
\begin{figure*}
\scalebox{0.7}{\includegraphics{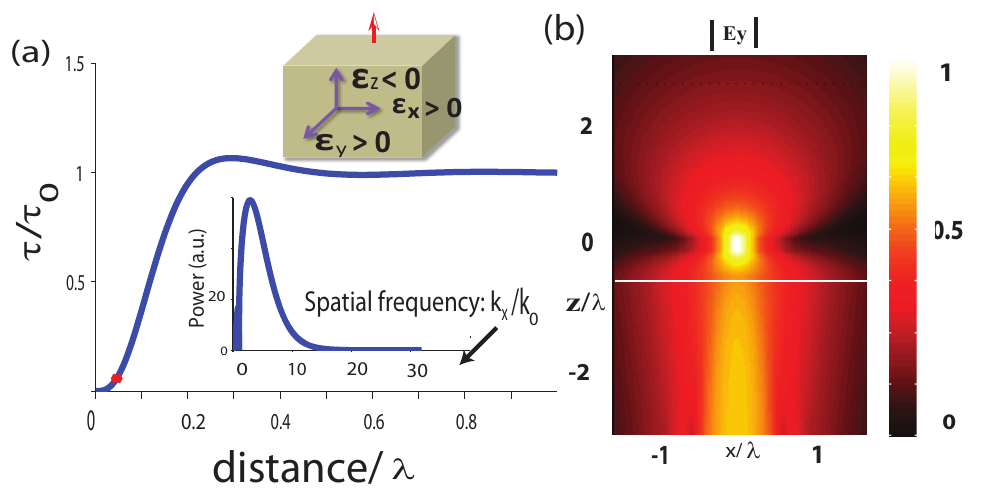}}
\caption{\label{fig2}{\bf a}) Spontaneous emission lifetime of a perpendicular dipole above a hyperbolic metamaterial substrate (see inset). Note the lifetime goes to zero in the close vicinity of the metamaterial as the photons are emitted nearly instantly. Most of the power emitted by the dipole is concentrated in the large spatial modes (evanescent in vacuum) which are converted to propagating waves within the metamaterial. (inset)  {\bf b}) False color plot of the field of the point dipole in a plane perpendicular to the metamaterial-vacuum interface (see inset of (a)) depicting the highly directional nature of the spontaneous emission (resonance cone).}
\end{figure*}
\clearpage
\begin{figure*}
\scalebox{0.5}{\includegraphics{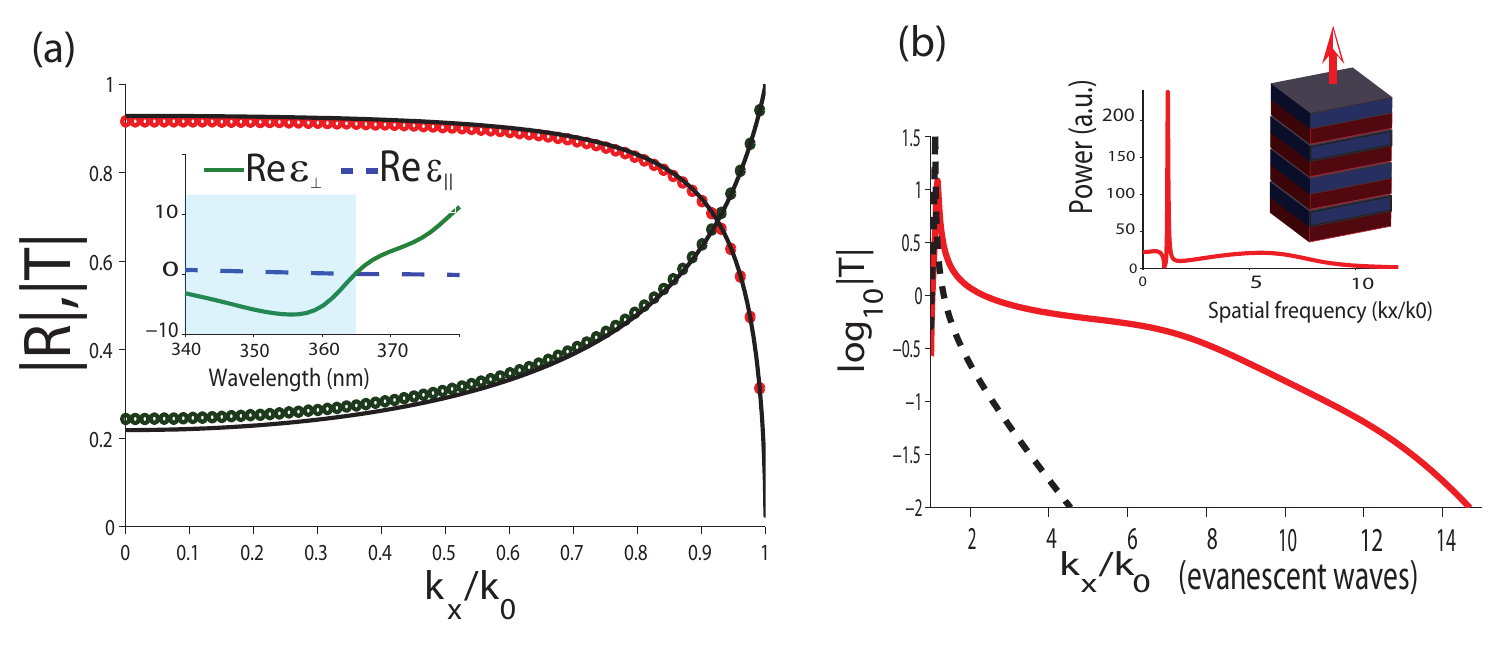}}
\caption{\label{fig3} {\bf a}) Comparison of the reflection and transmission amplitudes of plane waves incident on the planar multilayer realization of the hyperbolic metamaterial and effective medium theory. The metamaterial system consists of 8 alternating subwavelength layers of $Ag/Al_2O_3$. Green and red circles correspond to reflection and transmission computed using transfer matrix methods and the black superimposed line is calculated from effective medium theory. (inset) Hyperbolic dispersion is achieved in a broad bandwidth around $\lambda=350$ nm as the dielectric constants are of opposite signs in perpendicular directions. {\bf b}) Transmission of large wavevector waves by the layered hyperbolci metamaterial (red solid line) as compared to a high index dielectric (black dashed line). The inset shows the power spectrum of the dipole at a distance of $d=\lambda/20$ from the layered structure.  Most of the spontaneous emission occurs into high wavevector states which propagate within the h-MM}
\end{figure*}

\end{document}